\documentstyle[12pt,aaspp4]{article}

\begin{document}

\title{Galaxies at high redshifts --
observing galaxies in the cradle}
\author{E. Thommes}
\affil{Royal Observatory, Blackford Hill, Edinburgh, EH9 3HJ}
\authoremail{emt@roe.ac.uk}

{\bf Abstract \, \,}
Due to the invention of new powerful instruments in the
recent past (e.g. 10m class telescopes) high redshift galaxies are no 
longer a curiosity.\\
High redshift young star forming galaxies can be effectively discriminated
from the much more abundant foreground galaxies by their special
spectral properties: a prominent break at the Lyman limit
(i.e. a complete absence of flux at wavelength below the Lyman limit), 
an intrinsically flat spectrum at wavelength long-ward
of the Lyman limit and in the very early phase of evolution a
strong Ly-$\alpha $ emission line with high equivalent width.
In the last couple of years several hundred star forming galaxies
with $2.5 < z < 3.5$ could be identified using
deep broad band images which identify the Lyman break (Steidel et al.
1996, 1998). Spectroscopic followup observations confirmed their
high redshifts.
I summarize the main properties of these Lyman break galaxies (LBGs).
Furthermore, the very recent discoveries of very high-$z$ strong Ly-$\alpha $
emitting galaxies at $z>5$ and new large
systematic surveys (the Calar Alto Deep Imaging Survey and 
the survey of Hu et al. 1998) for such objects, suggest that the cosmic 
star formation rate in strong Ly-$\alpha $ emitters does not decrease at
redshifts $z > 3.5$ as suggested for the Lyman break galaxy sample.
I discuss the galaxies with the highest redshifts we know today and
give an overview over the survey of Hu et al. and the Calar Alto Deep Imaging
Survey.

\newpage



\section{Introduction}

The most obvious boundary conditions for our understanding 
of how galaxies formed come from the observable properties of 
the galaxies itself.

Due to the invention of efficient instruments and large
telescopes, e.g. large CCDs, NIR arrays, HST, Keck, etc. over
the last couple of years we now begin to record the evolution 
of galaxies from their first star formation at very high redshift.
In this context it is one of the great challenges of observational
cosmology to detect the ancestors of normal present day galaxies
(like our Milky Way) in the early universe which are in the stage
of their first star formation. In the following, 
I call such objects 
{\it primeval galaxies } (PGs).
The detection of a wide spread population of PGs would give us detailed
information about several aspects of galaxy formation and would
provide strong constraints for models of galaxy and
structure formation in the universe.

Recent spectacular detections of young galaxies at $z\ge 5$
(e.g. Dey et al. 1998, Weymann et al. 1998), the detection of a wide spread population
of young galaxies at $z=3...4$ using color selection methods (Steidel et al. 1996) and new large survey projects for Ly-$\alpha $ emitting PGs at redshifts
$z>5$ (the Calar Alto Deep Imaging Survey, Meisenheimer et al. 1998,
and the survey
of Hu et al. 1998), show that we are finally at the 
threshold where we are able to observe the earliest evolutionary phases
of galaxies -- galaxies in the cradle.

In this contribution I give an (biased) overview of the
recent progress in observing young galaxies out to $z\ge 5$.
The contribution is biased in the sense that I concentrate 
mainly on objects that we think could be the ancestors
of normal present day galaxies.

The impatient reader my leap straight to Fig. 13, where I present an overview
of a selection of 
individual objects and classes of objects with high redshift.
After a discussion of the expected 
spectral features of young star forming galaxies
at high-$z$ in section two, I discuss the main search techniques 
for high redshift galaxies in section three. In section four I
present the search technique, the status and main properties
of the Lyman break galaxies. In section five I discuss the search for
strong Ly-$\alpha $ emitting galaxies at redshifts $z>5$
and give final remarks in section six.

\section{Spectral features of young star forming galaxies at high z}

Although the nature of the faintest galaxies detectable 
on deep direct CCD-exposures is still one of astronomy's grand questions
(see the review of Ellis 1997),
it is well known that the majority of these objects are galaxies 
with redshifts below one. In order to distinguish between the bulk of these
``foreground'' galaxies and candidates for high-redshift galaxies ($z\ge 2$),
one has to find properties, which makes them different from the foreground
objects.
Therefore,
let me first summarize the expected spectral features of high-z young galaxies
which can be used
to identify them among the much more abundant foreground objects.
 
For redshifts $z>3$, optical telescopes record the rest frame UV
(1000 -- 2000 {\AA }). So one should look for interesting features
expected for young star forming galaxies especially in their rest-frame UV.
For every reasonable choice of the initial mass function
of the newly forming stars the (rest-frame) UV spectral energy distribution
of PGs should be dominated by the light of hot O and B stars. One could
expect that the spectrum may have similar features as the
UV spectra of giant HII regions in nearby galaxies.
Thus, by studying the spectra of nearby star-forming galaxies 
(see e.g. Heckmann \& Leitherer 1997) 
one should be able to learn something about
the expected properties of PGs at high redshift.
On the other hand, 
significant progress over the last years in modelling the 
evolution of stellar population in galaxies (see e.g. Charlot \& Fall 1993)
and the effects of the intervening intergalactic medium between us and
the high-z galaxies (Madau 1995),
allowed to calculate detailed models of the expected 
spectral energy distributions.
According to these models,
the most important features of the observable spectral energy
distribution of high redshift, young star forming galaxies are:

{\bf (i) Strong Ly-$\alpha $ emission}\\
Population synthesis models of Charlot \& Fall (1993) predict
that a young, dust free, star-forming galaxy (without an AGN)
should show strong Ly-$\alpha $ emission with (intrinsic) equivalent widths
in the range of 50-250 {\AA }.
The flux in the Ly-$\alpha $ line is expected to be directly proportional
to the actual SFR in the PG and 
3 -- 6 per cent of the bolometric luminosity are emitted in Ly-$\alpha $.
In the case of simple CASE B recombination and no dust, the Ly-$\alpha $ line
is expected to be 8--11 times stronger than H$\alpha $. 
On the other hand, because Ly-$\alpha$ photons are resonantly scattered,
the probability that Ly-$\alpha $ photons are absorbed by dust
is much higher than for continuum photons or photons of the Balmer lines.
A Ly-$\alpha $ photon in an HII-region will be scattered
10$^6$ to 10$^7$ times before it leaves the HII-region
(Osterbrock 1962). The optical path 
of a Ly-$\alpha $ photon to leave the HII-region 
is therefore $\sqrt(N) \approx 10^3$ times longer than the direct path.
Thus, even small amounts of dust in the plasma or in the surrounding
HI-gas of the HII-region are able to eat most of the Ly-$\alpha $ photons
away and could even lead to a negative Ly-$\alpha $ equivalent width (
Chen \& Neufeld 1994).
But the effect of interstellar dust on Ly-$\alpha $ photons
in connection with their resonant scattering is more complicated if
the interstellar medium consists of multiple phases.
For example, Neufeld (1991) showed that in a two phase interstellar medium
Ly-$\alpha $ photons could be less absorbed by dust
than continuum photons, resulting in even higher Ly-$\alpha $ equivalent
width than without dust.
However, in the very first phase of star formation in galaxies,
dust should not play a significant role. The presence of even small
amounts of dust is a signature that one generation of 
stars has formed already. Therefore, strong Ly-$\alpha $ emission could be a 
selection criterion, to select the youngest objects, which are just
in the stage of forming their first generation of stars, and which have not
yet enriched their interstellar medium with dust.
An additional factor which influences the detectability and especially
the line profile of the Ly-$\alpha $ line of PGs is the velocity
structure of neutral gas within the galaxies. As discussed above,
relatively small column densities of neutral gas with even a very small
dust content would destroy the Ly-$\alpha $ emission  if this gas is static
with respect to the ionized region where Ly-$\alpha $ photons originate from.
The situation changes when most of the neutral gas is
velocity-shifted with respect to the ionized regions (Kunth et al. 1998). 
For example, if the neutral gas surrounding the star-formation regions 
is outflowing from the ionized regions towards the observer, the 
resonant scattering would affect photons at shorter wavelengths than
the Ly-$\alpha $ emission line, allowing the Ly-$\alpha $
photons to escape at least partially. Energetic out-flowing interstellar
media are observed in the high redshift examples of Ly-$\alpha $ emitting
star-forming galaxies (see section 4) as well in local star-burst
galaxies observed with the Hubble Space Telescope (HST) and 
the Hopkins Ultraviolet Telescope (HUT) (Kunth et al. 1998; 
Gonz{\'a}lez Delgado et al. 1998).\\

{\bf (ii) An intrinsically flat spectrum ($F_\nu$) for $\lambda > 912$nm
(=Lyman limit)}\\
The second special spectral feature expected for young star forming
galaxies at high redshifts which is also predicted by 
population syntheses models (Charlot \& Fall 1993) is a remarkable
flat spectral energy distribution ($F_\nu \approx {\rm const} $)
between the Lyman-limit and the Balmer-limit.
This is due to  UV-radiation from hot, short-lived ($<10^8 {\rm yr}$) 
massive stars. In the case of a constant star formation rate (SFR)
an equilibrium between newly born and dying of these stars develops,
so that their number stays constant (which is directly proportional to the
SFR). Measuring the UV-flux of such a young star forming PG (which
is redshifted into the optical or near-infrared for $z>2$) 
is therefore a direct measure of the instantaneous SFR within the system.

{\bf (iii) A pronounced drop of the spectral energy distribution (SED) 
at the Lyman limit, e.g. a complete
absence of flux at wavelength below the Lyman limit ($\lambda < 912$nm) }\\
This ``Lyman break'' has three causes:
\begin{itemize}
\item[(a)] There is an intrinsic drop in the spectra of hot O and B stars at
the Ly-limit (Charlot \& Fall 1993, Cassinelli et al. 1995).
Furthermore, it is interesting that according to population syntheses models,
the spectra of a burst stellar population with
a declining star formation rate of time scale 10$^7$ yr develops
an extreme intrinsic Lyman break after about $10^8$ years 
(Charlot \& Fall 1993).
\item[(b)] Absorption by the neutral interstellar medium in the galaxy
itself (Leitherer et al. 1995a, 1995b, Gonz{\'a}lez Delgado et al. 1998)
\item[(c)] Photoelectric absorption from neutral hydrogen in the
Ly-$\alpha $ clouds and Ly-limit systems along the line of sight.
Madau (1995) computed the HI opacity of a clumpy universe as
a function of redshift, including scattering in resonant lines,
such as Ly-$\alpha $, Ly-$\beta$, Ly-$\gamma$, and higher order members,
and Lyman-continuum absorption. 
At wavelengths short-ward of Ly-$\alpha $ in the emitter rest frame but
long-ward of the Ly-limit, the
source's continuum intensity is attenuated by the combined
blanketing effect of many absorption lines of the Ly-forest
(Ly-$\alpha$, Ly-$\beta$, Ly-$\gamma $, Ly-$\delta $ ... line blanketing).
Photons with wavelength shorter than the Ly-limit in the emitter
rest frame suffer from photoelectric continuum absorption
from neutral hydrogen in the systems along the line of sight.
\end{itemize}

\section{Search techniques for high redshifted (primeval) galaxies}

The expected spectral properties of young star forming galaxies
at high redshifts discussed in the previous section lead to the following principle
techniques to detect them in the optical/near infrared window 
$0.32 \mu{\rm m} < \lambda < 2.4\mu{\rm m}$:
\begin{itemize}
\item[{\bf (A)}] {\bf Search for objects with prominent emission lines
with narrow-band filters.}
This requires the detection of emission line objects at a 
wavelength at which the redshifted lines (e.g. Ly-$\alpha $ or 
H$\alpha$) are expected. In order to keep the contrast with respect to the
night sky and continuum dominated objects high, one uses narrow band
imaging with a spectral resolution of a few hundred to several thousand 
km s$^{-1}$. Subsequent spectroscopic observations are necessary
to establish weather the detected emission line is in fact
the high redshift line one was searching for.
The highest redshift PGs which could be detected from the ground
by detecting their Ly-$\alpha $ emission line is given by redshifting this
line to the upper boundary of the K-band window
$\lambda_{obs}\approx 2.3 \mu$m which corresponds to a redshift of z=18.
\item[{\bf(B)}] {\bf Search for objects with unusual broad-band colours (using
the Lyman break as a criterion).}
This requires accurate broad-band photometry. The method was
successfully introduced by Steidel et al. (1992,1993, 1996a,b) and will
be discussed in section four.
\item[{\bf (C)}] {\bf A hybrid method, combining accurate 
broad-band and medium-band photometry with a narrow band search for line emitting objects.}
This allows one to get detailed information about the spectral energy distribution
of the objects, which can be used to discriminate between foreground
objects and good candidates for high redshift PGs. 
A survey based on this strategy is the Calar Alto Deep Imaging Survey (CADIS)
(see section 5.3).
\end{itemize}

The next decision one has to take for a search for high redshift object
is where to look for them.
Several strategies are used:
\begin{itemize}
\item[{\bf (i)}] {\bf ``Guided'' searches:}\\
The obvious targets for high-$z$ galaxy searches are fields in which
high-$z$ objects are already detected. According to the paradigm
of biased galaxy formation one would expect that the density
of high-$z$ galaxies should be enhanced in this fields.
Target fields for such ``guided'' searches have in the past been the fields
of high-$z$ QSO's, the 
fields of high-$z$ radio galaxies and searches at the redshifts
of damped Ly-$\alpha $ systems.
Ly-$\alpha $ emitting companions of quasars were indeed found
(Djorgovski et al. 1985, Hu et al. 1991, Petitjean et al. 1996).
Most of these are so close to the quasar, that their formation and 
excitation might be influenced by the quasar, so that they
may not be representative of normal galaxies. But there have also
been detections of high z Ly-$\alpha $ emitters in the fields
of high-$z$ quasars which are too far away from the quasars to be influenced
by the quasar itself  
(e.g. Hu and McMahon 1996).   
Damped Ly-$\alpha $ systems are thought to be signatures of disk
galaxies in the line-of-sight towards the quasar and possible clusters of galaxies at that redshift. The search for high-$z$ emission line galaxies
at the redshifts of damped Ly-$\alpha $ absorbers or strong metal absorbers
of QSO's were successful in several cases (e.g. Lowenthal et al. 1991,
Macchetto et al. 1993, Giavalisco et al. 1994, 
Moller \& Warren 1993, Djorgovski 1996, Francis et al. 1996, 
Mannucci et al. 1998).
\item[{\bf (ii)}] {\bf Search in ``empty'' fields:}\\
Fields suitable for the search of high redshift galaxies 
should not contain bright stars and should lie in a local minimum
of the IRAS 100 $\mu$m maps with e.g. absolute surface
brightness less than $2$ MJy/sr in order to avoid galactic extinction.
Over the last few years several systematic survey projects for 
high redshift galaxies in ``empty fields'' have started and 
have already proved to be very efficient
and successful. We will discuss some of these in section four and five.
\item[{\bf (iii)}] {\bf Systematic search for high redshift galaxies in
the surrounding of galaxy clusters.}
The magnification due to gravitational lensing enhances the
chance to detect high redshift objects (see e.g. Franx et al. 1997)
\item[{\bf (iv)}] {\bf Serendipitous discoveries}:\\
As always in science, some important discoveries are serendipitous.
For example, Dey et al. (1998) discovered a strong Ly-$\alpha $ emitting
galaxy at z=5.34 serendipitously during a successful search for $z\approx 4$ galaxies
using multi-colour techniques (see section 5.1).
\end{itemize}

\section{Lyman break galaxies}

\subsection{Search technique and status}

Based on the expected spectral property (iii) for high redshift
star forming galaxies (as described in section 2 and 3) 
that they should show an obvious
discontinuity in the far UV-spectrum at the limit of the Ly series 
near 912 {\AA} (the so called Lyman break), Steidel \& Hamilton
(1992, 1993) adopted a three filter system specifically tailored to
detect this Lyman break in objects which are at z$\approx 3$ .
Fig. 1 illustrates how the 3 filters sample the far - UV continuum of
a galaxy near $z\approx $3. Two of the filters ($U_n$ and $G$) have passbands
respectively below and above the Lyman-limit at $z\approx $ 3, while
the third filter, $R$, is further to the red.
As Fig. 2 shows, in deep images of the sky taken through these three
filters, $z\approx 3$ galaxies are clearly distinguished from the
bulk of lower redshift objects by their red ($U_n-G$) and blues
($G-R$) colours.
Follow up spectroscopy with the Keck telescopes confirmed that this
photometric selection technique is very efficient 
(Steidel et al. 1996a, 1996b).
Approximately 80 $\%$  of the robust candidates turned out to be
indeed high redshift galaxies, the remainder being mainly faint 
Galactic stars. Since then, large numbers of star forming galaxies at
$z\approx 3$ were discovered and confirmed using this technique
(see e.g. Pettini et al. 1997 for a review).
The Lyman break technique allows for the first time to assemble large
samples of galaxies at previously inaccessible redshifts in order to
study their properties and distribution.
Steidel et al. (1998b) compiled a large sample of $z\approx 3$ Lyman break
galaxies (LBGs) selected in a consistent manner in 5 survey, 
fields each of size 150-250 $\Box '$, which contain altogether
$\approx 1500$ LBG candidates (limiting magnitude $R\le 25$ mag).
They aim
to confirm $\approx 50\%$ of the candidates spectroscopically. 
Steidel et al. (1998b) claim to have 522 spectra of
galaxies with $z\ge 2.2 $, already (Mai 1998).
The multi-colour method to select high-$z$ galaxies was also
used in the Hubble Deep Field (HDF) and its flanking fields
(see Dickinson 1998 for a review).
The depth and the multi-colour nature of the HDF makes them ideal for
this purpose.

Using different filter systems, the original Lyman break selection
technique of Steidel et al. can be extended to higher redshifts.
Instead of looking for galaxies which exhibit the ``break'' in the $U$-band
(``$U$-dropouts'', $z\approx 3$) one can look for galaxies which
show the Lyman break in the $B$-band (``$B$-dropouts'') implying $z\approx 4$.
For example, Spinrad et al. (1998) used $BVRI$ images of the field around
the radio galaxy 6C0140+326 (z=4.41; Rawlings et al. 1996)
to select $B$-dropout candidates. They selected 13 potential $B$-dropout
candidates from which they already could spectroscopically confirm
6 to be at redshifts between z=3.6 and 4.02.
Steidel et al. (1998b) also used their $G$ and $R$ band filter images 
in connection with additional $i$-band images to select galaxies in 
the range $3.9\le z \le 4.5$ ($G$-dropouts).
They also got confirming spectra of galaxies in this redshift
range, although the sample is still much smaller than that for 
the $U$-dropouts.
An investigation of the $B$-dropouts in the HDF was done by Madau et al.
(1996). The remarkable result of their analysis is, that the space
density of Lyman break galaxies seems to be significantly
lower at $z\approx 4$ than at $z\approx 3$. Madau et al.
interpretated this as a declining cosmic star formation rate per comoving
volume for $z > 2.5 ... 3.5$. It will be important to check this
result with further observations in larger fields from the ground, although
the spectroscopic confirmation of galaxies at $z\ge 3.5$ gets
much harder because the spectral features useful for redshift
identification move to wavelength $\ge$ 6500 {\AA}, where the sky is
much brighter and the sky subtraction is more difficult.
However, very recently Weymann et al. (1998) were able to spectroscopically
confirm the redshift of a galaxy to be 5.60 which was selected as   
an ``$V$-band dropout'' in the HDF using the new NICMOS F110W ($\sim$ J)
and F160W ($\sim $H) data of the HDF together with the optical F606W ($V$)
and F814W ($I$) HDF data.
This shows that the multi-colour selection of high redshift objects
using the Lyman break as a signature can be extended to very high
redshifts, although the spectroscopic verification will get harder.

\subsection{Properties of Lyman break galaxies}

In the following I summarize the main observed properties of the
LBGs at redshifts 2.6$<$z$<$3.4 observed by Steidel et al.
(for details see Pettini et al. 1997, Steidel et al. 1998b, 
Dickinson 1998 and Adelberger et al. 1998).

{\bf (A) Surface density and comoving space density:}\\
The surface density of LBGs at $R = 25$mag  
which satisfy the colour criteria of Steidel et al. (see Fig. 1 and 2)
is 1.0/arcmin$^2$. This corresponds to a co-moving space density
of 2.2$\times 10^{-3}$h$^3_{70}$Mpc$^{-3}$ for $\Omega _m =1$,
which corresponds roughly to the present day space density of galaxies
with L$>$L$^*$.
For  $\Omega _m =0.2$ open or $\Omega _m =0.3$ flat (i.e. $\Omega _\Lambda = 0.7$) the space density
is roughly four times lower.

{\bf (B) Clustering properties:}\\ 
The redshift distribution of LBGs at z$\approx 3$ shows 
significant ``spikes'' corresponding to structures on a scale
of $\approx$ 10 Mpc (see Fig. 3).
Such structures would be extremely rare if galaxy number density fluctuations
were an unbiased tracer of matter fluctuations, and if one adopts a
``cluster normalisation'' for $\sigma _8$ (e.g. from Eke, Cole and
Frenk  1996). The existence of these ``high peaks'' requires
a significant bias of the galaxy fluctuations as compared to the
underlying mass fluctuations. From a count-in-cell measurement
of 268 LBGs with spectroscopic redshifts in six $9'\times 9'$
fields at $z\approx 3$ Adelberger et al. (1998) 
deduced a bias on scales of $\sim $ 10 Mpc of
$b= \delta _{gal} / \delta_{mass} = 6 \pm 1$, $1.9\pm 0.4$ and
$4.0\pm 0.7$ for $\Omega _m =1$, $\Omega _m =0.2$ open, and
$\Omega _m = 0.3$ flat respectively.

Both the number density (which reflect the power on
scales of $\sim $ 1 Mpc) and the clustering properties
(which reflect the power on $\sim $ 10 Mpc scales)
together give constraints on the shape of the power spectrum
of dark matter models of structure formation. According to these
models, the large bias values are understandable, if individual
LBGs would be associated with dark haloes with virial velocities of
$\approx 150 {\rm km/s}$,
similar to the dark haloes of massive galaxies at the present epoch.

{\bf (C) Spectra of LBGs:}\\
Fig. 4 shows some examples of spectra of LBGs.
They show many similarities with those of nearby star-burst galaxies.
A typical z$\approx 3$ LBG with $R\approx 24.5$, $(G-R)\approx 0.5$
has a far UV luminosity of 
$L_{1500} \approx 1.3\times 10^{34} h^{-2}_{70} $ W/{\AA}.
This corresponds to a star formation
rate (SFR) of $\approx 8$ h$^{-2}_{70}$ M$_\odot$/yr
(assuming a continuous star formation model with an age greater
than $10^8$ years and a Salpeter IMF from 0.1 to 100 $M_\odot$
, see e.g. Leitherer et al. 1995, $H_0 = 70 h_{70} {\rm km s^{-1} Mpc^{-1}}$.) 
The spectra show strong low-ionization interstellar absorption lines e.g.
SiII$\lambda$1260{\AA}, OI$\lambda$1302{\AA}, CII$\lambda$1335{\AA},
FeII$\lambda$1608{\AA}, AlII$\lambda$1670{\AA} as well as high-ionization
stellar absorption
lines, often with P-Cygni line profiles characteristic of
Wolf-Rayet and O star winds.
Only 75$\%$ of the LBGs show Ly-$\alpha $ emission and always weaker
than expected on the basis of the UV continuum luminosities.
Some LBGs show Ly-$\alpha $ even in absorption.
There are hints of a correlation between the strength of the
metallic interstellar medium absorption and the strength of Ly-$\alpha $
absorption, and an anti-correlation with Ly-$\alpha $ emission (Spinrad et al.
1998).
In the cases in which Ly-$\alpha $ is detected, the line is generally
redshifted up to several hundred km/s relative to the interstellar
absorption lines, and its profiles are clearly asymmetric,
showing a P-Cygni type profile.
This can be understood as originating from resonant scattering of
the Ly-$\alpha $ photons in an out-flowing interstellar medium driven
by mechanical energy generated in the star formation episode.
As already mentioned above, similar effects are observed in nearby 
star forming galaxies. 

{\bf (D) Dust content of LBGs:}\\
The UV spectral slope $\alpha $ of LBGs ($f_\nu \sim \lambda ^{\alpha }$)
are typically between 0 and +1.5 and therefore significantly
greater than expected from spectral models which predict
$\alpha \approx -0.5  ... 0$. This can be explained if the spectra
are reddened by dust extinction. Recently Pettini et al. (1998)
presented the first five infrared spectra of LBGs in which they successfully
detected Balmer and [OIII] emission lines. The detected H$\beta $ luminosities
(uncorrected for dust extinction) implied star formation rates of 
20-270$ h^{-2}_{70}{\rm M_{\odot} yr^{-1}}$ 
which are significantly greater than those deduced from 
the UV luminosities at 1500 {\AA} 
(see above). Although the sample is still very limited and 
uncertainties in the shape of the reddening curve and in the intrinsic
UV continuum slope does not allow yet to determine the level of the dust 
extinction accurately, Pettini et al. estimated that an extinction of
1-2 magnitudes at 1500 {\AA} may be typical for LBGs.

{\bf (E) Speculations about the mass of LBGs}\\
In four of the five galaxies for which Pettini et al. (1998) 
detected Balmer and [OIII] emission lines, they measured a velocity
dispersion for the emission line gas of $\sigma \approx 70$km/s.
With typical half-light radii of $\approx 2$kpc (deduced from HST 
NICMOS images as well as high resolution Keck images) Pettini et al.
suggested virial masses for the 
LBGs in the range of M$_{\rm vir}\approx 1-5\times 10^{10}$ M$_{\odot}$
in contrast to dark halo masses of  $M_{\rm DM} \ge 10^{11}{\rm M}_\odot$ 
suggested
from the number density and clustering properties (see above).
But as Pettini et al. pointed out, it is possible that
velocity dispersions (and therefore the masses deduced from them)
have been significantly underestimated because the current limited 
sensitivity of the IR-spectra may only sample the inner cores of the
galaxies, where the star formation rate is high.

\subsection{What are the LBGs ?}

For a detailed discussion of the interpretation of LBGs see
the contribution of Mo in this proceedings.
I want to mention here only two points:

(i) The number density and clustering properties of LBGs are consistent
in all currently popular hierarchical models, if they are the central
galaxies of the most massive dark matter halos present at $z\sim 3$.

(ii) The LBGs are not the long searched for primeval galaxies.
They already formed at least one generation of stars which chemically
enriched the systems and lead to dust formation, making Ly-$\alpha $
emission weak or undetectable.
Earlier stages of evolution, before the formation of dust, may have much
stronger Ly-$\alpha $ emission and fainter stellar continuua than
the LBGs at z$\approx 2.5 ... 3.5$ and may therefore be hard to pick
out with the colour-break techniques, but relatively easy to find by
searches for emission line objects.
This is the subject of the next section.

\section{Search for Ly-$\alpha $ emitting high redshift galaxies}

The systematic search for Ly-$\alpha $ emitting high redshift PGs has
for a long time been a frustrating business without the success
expected from early predictions (for a review see e.g. Pritchet 1994).
But Thommes \& Meisenheimer (1995, 1998, see also section 5.4 below)
showed, that these searches were all based on far too optimistic
predictions and that even the deepest surveys up to now had no realistic
chance to find Ly-$\alpha $ emitting objects at high redshifts 
in large numbers.

Recently, a couple of strong Ly-$\alpha $ emitting galaxies at 
redshifts $\ge 5$ have been discovered, and two new large survey projects
which are much deeper and/or cover larger areas on the sky 
aiming to discover Ly-$\alpha $ emitters at redshifts $z=3...6$
have been started.
One of these two surveys is the Calar Alto Deep Imaging Survey (CADIS)
of the Max-Planck-Insitut f\"ur Astronomie in Heidelberg, Germany
(Meisenheimer et al. 1998, Thommes et al. 1997, 1998a).
The second 
one is a survey of Hu \& Cowie (1998) at the Institut for Astronomy in 
Hawaii.
First results of these surveys, and the already mentioned
recent discoveries of individual strong Ly-$\alpha $ emitters at
redshifts of 4.92, 5.34 and 5.6, suggest that there might indeed
exist an abundant population of strong Ly-$\alpha $ emitters at
$z\ge 5$ which might have a more ``primeval signature'' than the LBGs.

\subsection{Examples of high redshift Ly-$\alpha $ emitting galaxies}

As I mentioned in section 3, targeted searches for strong Ly-$\alpha $
emitters at the
redshifts of quasars/radiogalaxies and the redshifts of absorption systems
detected in quasar spectra, had been successful in several cases
(e.g. Lowenthal et al. 1991,
Macchetto et al. 1993, Giavalisco et al. 1994, 
Moller \& Warren 1993, Djorgovski 1996, Francis et al. 1996, 
Pascarelle 1996, Hu\& McMahon 1996).
In 1997 Franx et al. reported the serendipitous discovery of two
strong Ly-$\alpha $ emitting gravitationally lensed galaxies
which they could spectroscopically confirm to be at a redshift of
z=4.92. At that time these galaxies were the objects with the highest
redshifts and for the first time the redshift record was not held
by a quasar (the highest quasar redshift is still 4.897).
One of the two galaxies is a prominent red arc visible in the
HST images of the cluster CL1358+62. Franx et al. obtained Keck spectra
of the arc, showing strong Ly-$\alpha $ emission at 7204 {\AA},
a continuum drop blue-ward of the line and several absorption lines
to the red. They reconstructed the image of the high redshift galaxy using
a gravitational lens model. The reconstructed galaxy image is
asymmetric, containing a bright knot and a patch of extended emission
0.4 arcsec from the knot. This irregular structure is in contrast with
the compact morphologies of the LBGs at $z\approx 3$ (Giavalisco et al. 1996).
The galaxy has a luminosity I$_{AB}\approx 24$ mag, 
which could be produced by a star formation rate of 18 h$^{-2}_70$
M$_{\odot}$yr$^{-1}$ (($q_0=0.5$). The spectral lines show velocity
variations on the order of 300 km/s along the arc.
The Ly-$\alpha $ emission line is asymmetric with a red tail and the
Si V line is blue shifted with respect to Ly-$\alpha $. These characteristic
features already well known from the LBGs (see above) again indicate
that radial outflows dominate the kinematics
of the absorption line gas.
The second galaxy Franx et al. confirmed to be at $z=4.92$ is a companion
galaxy with a radial velocity of only 450 kms$^{-1}$ different than
that of the arc.
Although these two galaxies with $z=4.92$ were the first objects
which beat the quasars in redshift, they were soon beaten by an other
serendipitous discovery of a strong Ly-$\alpha $ emitter at z=5.34 by
Dey et al. (1998). This galaxy was the first object spectroscopically
confirmed to be at z$>$5 and was discovered during a successful search
for z$\sim$ 4 LBGs ($B$-dropouts) (see Fig. 5).
Fig. 6 shows the spectrum of this z=5.34 Ly-$\alpha $ emitter,
which is exposed 36200 s with the Low Resolution Imaging Spectrometer
(LRIS; Oke et al., 1995) on the Keck telescope.
The strong emission line is asymmetric, as expected for Ly-$\alpha $ lines
in high redshifted young star forming galaxies, and the spectrum shows
a weak continuum emission (= 27AB mag for 7735 {\AA} $<\lambda _{obs} <$
9000 {\AA}) red-ward of the line and no flux detection (upper limit
=29.5AB mag 1$\sigma$) blue-ward of the line (6200 {\AA} $<\lambda _{obs} <$
7700 {\AA}).
Dey et al. measured a ``flux-deficit'' parameter 
\footnote{which is defined as $D_A = < 1 - {F_\nu(\lambda 1050-1170)_{obs}
\over {F_\nu(\lambda 1050-1170)_{pred}}} > $.$D_A$ describes the discontinuity
at the Ly-$\alpha$ line due to absorption in foreground intervening systems.
The predicted flux is the flux level one would expect without this absorption.}
for this
discontinuity at the emission line of D$_{A}>0.70$
(3$\sigma $ lower limit) which is consistent with the the
theoretical estimate of D$_A$(z=5.34)=0.79 of Madau (1995) for a galaxy
at z=5.34.
Furthermore, because there are no other lines detected in the spectrum,
there are just two possible interpretations for the emission line:
[OII]$\lambda \lambda$3726,3729  (z=1.07) or Ly-$\alpha $ (z=5.34).
In case of [OII] at z=1.07, the rest frame equivalent width would be
$\approx $ 300 {\AA}, which is in excess of what is produced by even 
the most luminous star-forming galaxies (typical less than 70 {\AA},
Liu \& Kennicutt, 1995). In addition, Dey et al. tried to fit the 
observed line profile with an [OII]$\lambda \lambda$3726,3729 
doublet which resulted in a 3726/3729 ratio of $>$2 which is
unphysical (the highest allowed is 1.5).
Thus, the only interpretation for the emission line is indeed Ly-$\alpha $ and
therefore the redshift of z=5.34 is secure.
The relatively narrow Ly-$\alpha $ line (deconvolved FWHM $\approx $
280 kms$^{-1}$), the rest frame Ly-$\alpha $ equivalent width of 
$\approx 95 \pm 15${\AA}, and the lack of strong emission of 
NV $\lambda \lambda$1239,1243 and SiIV $\lambda \lambda $1394,1403,
suggest that this is a star-forming galaxy and not an AGN. 
The star-formation rates deduced from the UV continuum emission and
the Ly-$\alpha $ emission coincides and gives a SFR$\approx 3$M$_\odot$/yr
h$^{-2}_{70}$ (for $q_0=0.5$).
Although the deduced SFR is comparable to the SFR observed in the LBGs,
the fact that the SFR implied by L$_{1500}$ is very similar to that
derived from L$_{Ly\alpha}$ shows, that this object is not
significantly attenuated by dust, in contrast to the LBGs (see above).
Furthermore, the galaxy at z=5.34 discovered by Dey et al. is spatially
resolved in the ground-based I-band images. Although the observed 
morphology may be dictated by the Ly-$\alpha $ emission line
(which is contained in the I-band filter), this is again differing
from the LBGs at redshifts 2.5--3.5, which are compact systems with 
half-light radii $\sim$ 0.2--0.3 arcsec
(Giavalisco et al. 1996).
These differences could be understood, if the galaxy
at z=5.34 is in an earlier stage of evolution, than the LBGs.

At the time of the writing of this contribution (October 1998)
the redshift record has already been pushed beyond z=5.34.
As I mentioned above, Weymann et al. (1998) selected a galaxy
on the new HST NICMOS J and H images of the HDF which could be spectroscopically
confirmed to be a strong Ly-$\alpha $ emitter at z=5.60.
Furthermore, Hu et al. (1998) claim to have found a strong
Ly-$\alpha $ emitter at a redshift of z=5.64 in their survey (see below).

\subsection{Blank-field survey for high-z Ly-$\alpha $ emitters
of Hu et al. (1998)}

Using the 10m Keck II telescope, Hu, Cowie \& McMahon (1998)
recently started a narrowband  and spectroscopic search for 
strong Ly-$\alpha $ emitting star forming galaxies in the redshift
range 3--6. They reach extremely faint flux levels of
$\approx 1.5\times 10^{-20}$ W/m$^2$ (5$\sigma $). 
Their main selection criterion to discriminate high redshift
Ly-$\alpha $ candidates from foreground emission line objects 
is an extremely high equivalent width of the emission line ($W_\lambda
> 100$ {\AA}) together
with additional broad-band colour criteria.
As a first step, they performed deep narrow band images with a 5390/77 {\AA }
filter together with deep broad band $B$,$V$ and $I$ data in the HDF and the
Hawaii Deep Field SSA22 with the LRIS instrument at the Keck II
10m telescope (see Cowie et al. 1998). Using the redshift
information available for many objects in the HDF and SSA22, they
demonstrated how certain criteria on the $(V-I)$ and $(B-V)$ colour
can help to discriminate between the three main classes of
strong emission lines [OII], [OIII] and Ly-$\alpha $, that can produce
significant equivalent widths in the narrow band.
Using these colour criteria together with the equivalent width
criterion (W$_{\lambda}>100${\AA } in the 5390/77 {\AA} filter),
 they selected a sample of 12 objects in the SSA22 and HDF fields.
Among these 12 objects is the Ly-$\alpha $ emitting object, whose
redshift was already confirmed to be at $z = 3.4$ by
Lowenthal et al. (1997).
Multi-object spectroscopy of the objects with LRIS on Keck II
confirmed the strong emission lines to be Ly-$\alpha$. 
Some of these Ly-$\alpha $ emitters show continuum colours similar
to those of the colour-selected LBGs (which is not surprising).
However, the most interesting result is that the sample also contains
objects with very faint continuua, that would have fallen below the
magnitude threshold of current LBG surveys. Two of the objects 
could not even be detected in the broad band images at all
(1$\sigma$, $B$=27.8, $V$=27.5, and $I$=25.8) with W$_\lambda >$ 400 {\AA}.
Although the covered field of 46 arcmin$^2$ and redshift range of
3.405-3.470 is very small, Cowie and Hu (1998) deduce from their detections
that the number density of strong Ly-$\alpha $ emitters at $z\approx 3.5$ with
flux $>2 \times 10^{-20}$W/m$^2$ could be as high as 13000/unit z/ 
$\Box ^\circ$ of which an essential fraction has too faint continuua
to be selected by current LBG surveys.

Hu, Cowie \& McMahon extended their search for Ly-$\alpha $ emitting
galaxies at high z by observing the SSA22 field through a narrow-band
interference filter centered at 6741 {\AA} with a bandpass of 78 {\AA},
corresponding to a Ly-$\alpha $ redshift of z$\approx $4.52.
They selected 3 emission line objects with observed equivalent width
of $W_{\lambda } > 100${\AA}. To discriminate between the most likely
interpretations of the emission line
([OII]$\lambda $3727 at $z=0.$8, [OIII]$\lambda \lambda$4959,5007 or
H$\beta$ $\lambda $4861 at z=0.37, H$\alpha $ $\lambda $6563 at $z=0.03$ 
or Ly-$\alpha $ at $z=4.52$), 
Hu et al. did spectroscopic follow up observations
of all the candidates. The lowest flux object with the weakest
equivalent width ($W_\lambda \approx $ 120 {\AA}) could be identified as
an [OII] emitter at z=0.814 and the other two are indeed most likely
Ly-$\alpha $ emitters at $z=4.52$

In addition to these narrow band searches, Hu et al. examined
older spectroscopic data taken for other purposes, in order to search
for serendipitous high redshift Ly-$\alpha $ emitters.   
The data (slit spectra with LIRS on the 10m Keck telescope) covered
a blank sky area of $\approx 200 \Box '$, covered a spectral range
corresponding to Ly-$\alpha $ redshifts of z=3.04-5.64, and
reached a detection limit of 1$\sigma \sim 10^{-21}$W/m$^2$s$^{-1}$. 
In this data set they found 4 high equivalent width objects which show hints
of a break at the line. If the lines they detect are indeed Ly-$\alpha$,
the redshifts would be $z=5.64, 4.19, 4.02, 3.05$.
However, especially the spectrum of the probably $z=5.64$ object
is not good enough to be completely sure that the detected line is
Ly-$\alpha$. The spectra do not have
the necessary S/N to see e.g. the expected asymmetry of a 
high z Ly-$\alpha $ line, to see the doublet structure in case of [OII],
or to measure the flux decrement $D_A$, like it was possible in the spectrum
of the z=5.34 Ly-$\alpha $ emitter discovered by Dey et al. (1998).
Additional spectroscopic data is necessary.

However, this first results of the Hu et al. survey show that
Ly-$\alpha $ emitting galaxies at high redshifts are quite common objects
with a significant contribution to the cosmic star formation rate.
Although the covered area is still much too small to draw conclusions
there are already hints that the cosmic star formation rate in the
strong Ly-$\alpha $ emitters is
 probably
constant or even increasing with redshift from $z=3-6$, in
contrast to colour-based LBG samples where the rate seems to be 
declining at higher redshifts (Madau et al. 1996, 1998).
However, this has to be proven by additional data.
We started a large emission-line survey 
for high-z Ly-$\alpha $ emitting primeval galaxies (z=4.7, 5.7, 6.5)
at the Max-Planck Institut f\"ur Astronomie in
Heidelberg, called the Calar Alto Deep Imaging Survey (CADIS), which covers a much larger area on the sky 
($\approx 0.3 \Box ^\circ $) reaching flux limits of
$3\times 10^{-20}$ Wm$^{-2}$. 
With its large area coverage CADIS complements the deeper Hawaii survey
and both together will give interesting constraints for the number density
of Ly-$\alpha $ emitting primeval galaxies at z$>$5.  

\subsection{The Calar Alto Deep Imaging Survey}

\subsubsection{The CADIS concept}
  
The Calar Alto Deep Imaging Survey (CADIS) is the current extragalactic
key-project of the Max-Planck-Institut f\"ur Astronomie (see Meisenheimer et al. 1995,1998, Thommes et al. 1997)
\footnote{Participating scientists: S. Beckwith, 
F.H. Chaffee, R. Fockenbrock, J. Fried, H. Hippelein, U. Hopp,
J. Huang, Ch. Leinert, K. Meisenheimer, S. Phleps, H.-J. R\"oser, I. Thiering, 
E. Thommes, D. Thompson,
B. von Kuhlmann, C. Wolf}.
CADIS is a very deep emission line survey using an imaging Fabry-P\'erot, 
combined
with deep broad- and medium-band photometry. This survey project
is specifically designed to detect Ly-$\alpha $ emitting 
primeval galaxies (PGs) at redshifts $z\ge 5$, but
it will in addition produce a large data base for investigations of
the faint end of the 
luminosity function and the three dimensional correlation function of
faint emission line galaxies at intermediate redshifts ($0.2 < z < 1.2$).
Its multifilter technique also allows the classification and redshift
determination of several hundred early type galaxies at 0.5 $<z<1.2$
(which will complete the luminosity function at the bright end), 
to detect faint QSOs beyond $z=3$ and even beyond $z=5$, 
to select extremely red objects (EROs) with $R-K' \ge 6$, and of faint M stars
and faint brown dwarfs.

The central point of the strategy is the search for faint emission
line objects ($S_{lim}(5\sigma ) \approx 3\times 10^{-20}$W/m$^2$)
with an imaging Fabry-P{\'e}rot in the wavelength intervals 
(A) [694,706nm], (B) [814,826nm] and (C) [910,926nm]. 
These wavelength intervals are located in prominent windows
of the night sky emission, corresponding to Ly-$\alpha $ redshifts of 
z=4.75, 5.75 and 6.55. 
The very deep flux limit for line detection
allows one to detect star formation rates (SFRs) of 10 to 50
M$_\odot$/yr in PGs at $z\ge 5$.
A completely new feature of the CADIS strategy are very deep images in
a set of narrow band filters ($R = \lambda / \Delta \lambda \approx 50$,
typical exposure times: 10 h at a 2.2 m telescope), which are selected
in such a way, that for every prominent emission line
of a foreground object falling into the Fabry-P{\'e}rot scan
(e.g. $H\alpha $) a second or third line (e.g. [OII]372.7nm or
[OIII]500.7nm ) should show up in them (fig. 7).
We call these narrow band filters veto-filters since a signal in one
of them excludes that the line detected in the Fabry-P{\'e}rot is
Ly-$\alpha $. Combining the veto-filter information with the accurate
spectroscopy through the Fabry-P{\'e}rot, we expect to determine the
redshift of the majority of emission line galaxies ($0.25 < z < 1.0$)
to an accuracy of 100 ... 200 km/s (depending on the S/N ratio).
Since the detection of emission line objects requires several ``off-line''
continuum exposures anyway, we decided to add in a complete set of
broad band and medium band images, which are optimized both for the
continuum determination  and the identification of one of the most
severe contaminations - faint M-stars in our Galaxy.
The optical multi-color survey is supplemented by a $K'$
(5$\sigma $ limit 21.0 mag) survey with the new OMEGA camera at the
prime focus of the Calar Alto 3.5 m-telescope (6.6 $\times $ 6.6 $\Box '$
field). This gives us a global view of the spectral energy distribution (SED)
of every object in the field.
The narrow band and medium band images will allow us to 
discriminate very effectively between foreground emission line objects
and good candidates for Ly-$\alpha $ emitting PGs at high redshifts
(Fig. 7).
CADIS will survey 10 fields (each 120 $\Box '$) distributed over the
northern sky ($\delta \ge -5^{\circ}$), which were selected for their
absence of bright stars ($R\le 16.0$) and an extra-ordinary low flux in
the IRAS 100$\mu$m maps ($\le$ 2 MJy/sr). Thus, the total survey area
will be at least 0.3 $\Box ^\circ $. CADIS will use the 2.2m- and the 
3.5m-telescopes at Calar Alto (Spain).

\subsubsection{Current status of the survey}
About 1/3 of the survey data have been collected in the two years
since regular survey observations commenced at the Calar Alto 2.2m and
3.5m telescopes. Delays in the delivery of the Fabry-P{\'e}rot-etalon
for the 3.5m focal reducer, however, have caused a large backlog in the 
central part of CADIS - the deep emission line survey. Up to now, we have only
been able to collect reasonably complete data sets for two northern
fields at RA=09h and 16h.

\subsubsection{Primeval galaxies at $z>5$ - problems and first candidates}

Due to the delay in the delivery of the Fabry-P{\'e}rot etalon 
for the 3.5m focal reducer the emission line survey is far behind our
original schedule.
In addition, we recently discovered that the original flat field corrections
of the Fabry-P{\'e}rot images were severely flawed by internal reflections 
between the etalon and the focal reducer optics, which causes
a central ``hump'' in the images.
Treating this central ``hump'' as a (multiplicative) flat
field property thus artificially enhanced the measured flux
in the Fabry-P{\'e}rot bands in the outer regions by $> 20 \%$.
We subsequently developed a method by which an additional calibration
through a multi-hole mask inserted at the telescope focus is used to calibrate
the transmission properties of the entire optical system 
focal reducer-etalon-prefilter.
The central hump (which is also present on night sky science exposures)
is then treated as sky concentration and subtracted from the science exposures.
Previously, we presented a list of seven Ly-$\alpha $ candidates, drawn from an early
analysis of first CADIS observations in the 9h field (Thommes et al. 1998a).
However, a reanalysis of the data with the improved flat field correction,
together with new Fabry-P{\'e}rot and veto-filter observations as well
as spectroscopic follow-up observations of some of the candidates with
the Keck 10m telescope, showed that none of these candidates is 
a genuine Ly-$\alpha $ galaxy at redshift $z=5.7$ (for details see
Thommes et al. 1998b). One of the candidates seems to be an emission
line object which is located a $z\approx 0.25$. The remaining 6 objects
are most likely instrumental artifacts which have been 
caused by the imperfect flat field correction of the Fabry-P{\'e}rot 
data in the first analysis.
There seems to be no emission line objects detected in the 9h field,
which cannot be identified as a foreground galaxy by means of the
veto filter observations or a significant signal on the deep blue
band exposures. 
That the CADIS Fabry-P{\'e}rot and narrow band data are indeed
able to identify the emission lines and to determine the redshift
of the faint foreground emission line objects, is demonstrated in 
Fig. 8.

Applying the improved data reduction procedure to the - much more 
complete - Fabry-P{\'e}rot and veto filter data of the 16h field
yields two Ly-$\alpha $ candidates, one of which appears uncertain
due to its proximity to a rather bright stellar object (R=19.6).
In summary, we are currently left with 1$\pm 1$ Ly-$\alpha $ candidate
at $z=5.7$ in a search volume corresponding to one complete
CADIS field (7 wavelength settings in 16h field and 3 settings in the
9h field in wavelength region B [814,826nm]).

\subsubsection{Surface density of Ly-$\alpha $ emitting galaxies at
high redshifts}

Assuming that there are in any case less than 3 Ly-$\alpha $ emitting
PGs at z=5.7 in our 9H and 16H field data, we get the limit marked
in Fig. 9. For comparison, the small points with
the error bars correspond to the candidates reported by
Cowie and Hu (1998) for Ly-$\alpha $ emitters at z=3.4.
If the detection of the very high surface density
of Ly-$\alpha $ emitting galaxies by Cowie and Hu (1998) and Hu et al. (1998)
at z=3.4 and 4.5 as shown in Fig. 9 will be confirmed, the limits posed by
our non-detection of PGs in the small subset of CADIS data may give challenging constraints for models of the surface density of Ly-$\alpha $ emitting
galaxies at high redshift. 
The curves in Fig. 9 show the results of such model calculations.
We trimmed the free parameters (see Thommes \& Meisenheimer, 1998c)
to get results which
are consistent with our limits and the observations of Cowie and Hu (1998).
The models predict that the number density of
Ly-$\alpha $ galaxies should not change very much in the redshift interval
$z_0=3.5 $ to 4.5 at low detection limits of $S_{lim} \le 2\times 10^{-20}$ W/m$^2$.
This is consistent with the claims of Hu et al. (1998), that the surface density
of Ly-$\alpha $ emitters at these detection limits is  as high
as $\approx 1500/\Box ^\circ / \Delta z=0.1$ at $z_0=3.5$ {\bf and} 4.5.
At detection limits above $S_{lim} \ge 2\times 10^{-20} $ W/m$^2$
the expected number density decreases rapidly with increasing $z_0$,
consistent with 
the null detection in the recent subset of CADIS data.
At redshifts $\ge 6.0$, the models predict 
surface densities which are much lower than at $z_0=3.5$ for all 
detection limits $S_{lim} \ge 10^{-20} $ W/m$^2$.
However, much more data in more fields is needed to draw definite conclusions
about the mean surface density of Ly-$\alpha $ emitting galaxies at high redshifts. Especially, if Ly-$\alpha $ galaxies at $z \ge 3.5$ have similar clustering
properties as the colour selected Lyman break galaxies, the number density
of Ly-$\alpha $ emitters in the CADIS data and the data of Cowie and Hu
may be substantially influenced by clustering.
Steidel et al. (1998) found high spikes with relative overdensities of up to
2.6 in the distribution of Lyman break galaxies in volumes of comoving size
$\approx 10^3$ h$^{-3}_{100}$ Mpc$^3$ (see section 4.2). The 9H CADIS data cover a
comoving volume of $27.1\times 27.1 \times 12.7 h^{-3}_{70}$ Mpc$^3$ in the case of $q_0 = 0.1$ and
$14.6\times 14.6\times 7.1 h^{-3}_{70}$ Mpc$^3$ in the case of $q_0 = 0.5$. Thus, clustering
may substantially influence the number density of high-z Ly-$\alpha $ emitters
observed in single CADIS fields. 
To average these effects out, one has to observe many fields and/or larger 
redshift intervals.
CADIS will search for Ly-$\alpha $ emitting galaxies in 
9 fields each of $\approx $ 120 $\Box '$ at redshifts 4.75, 5.75 and 6.55 with
coverage of $\Delta z = 0.1$ each. This should give a firm basis to investigate
the average number density of Ly-$\alpha $ emitting galaxies at high redshifts and
may produce strong constraints for models of galaxy formation.

\subsubsection{Other interesting first CADIS results: High-z QSOs and EROs}

CADIS identifies QSOs either by their peculiar colours or their variability
over a period of several years. The complete optical filter set of the survey
is displayed in Fig. 10. 

CADIS employs a completely new
approach for classification and redshift estimation from 
multi-colour data. Essentially it is based on a comparison between
the {\it observed} colours of any objects detected in the survey with the
``reference colours'' of tens of thousands of template spectra of stars,
galaxies and QSOs of different type and redshift (for details see Wolf, C.
et al. 1998). Applying this method to the data of three CADIS fields 
of 225 $\Box' $ total area, we selected 26 objects with AGN-like colours.
Spectroscopic followup observations confirmed 22 of the 26 candidates
as broad-line AGNs, whereas three turned out to be stars and one is
a narrow emission-line galaxy.
Five of the 22 AGNs have luminosities below $M_B=-23$mag ranking them
as Seyfert-1 galaxies, while the other 17 are genuine quasars. Among
these, 16 are brighter than $R=22$ mag  and half of them are located 
at $z>2.2$. Our results for the surface density of $z<2.2$-quasars
are consistent with published values, but at $z>2.2$ the CADIS number
exceeds published values (see Fig. 11).
This is a very interesting result, which challenges the
general belief that the quasar number density drops off at higher redshift,
as suggested by previous surveys (see review of Hartwick \&
Schade, 1990).
Obviously, CADIS finds high-redshift quasars with unprecedented completeness.
An explanation for the fact 
that other surveys miss many of the quasars with redshifts higher than
2.2, which CADIS
is able to find, is that many of these quasars found in CADIS display
rather star-like colours in broadband filters, and would not have been
picked out in classical multi-colour surveys.

Another interesting result of CADIS is the detection of many extremely
red objects (EROs,) defined by their $R-K'$ $\ge $ 6. To date we  
have observed 4 CADIS fields, each covering $\approx 160 $ arcmin$^2$
to a 5$\sigma$ limit of $K'$=20.5 mag. Combining these data with
the CADIS $R$-band data (2$\sigma$-limit =26.0 mag), we have selected
complete, unbiased, magnitude-limited samples of EROs (see Fig. 12). 
We find a surface density of EROs with $K'\le 20.0$ mag of 0.33 arcmin$^2$ and
for EROs with $K'\le 19.0$ mag of 0.039 armin$^2$ (see Thompson et al. 1998).
Two alternative interpretations have emerged to explain the extremely
red spectral energy distributions (SEDs) of these galaxies.
Initially, an old stellar population, such as found in present-day
elliptical galaxies, provided a good fit to the SEDs. A strong,
redshifted 4000 {\AA} break falling between the $R$ and $K$ 
filter bandpasses and
the lack of any appreciable restframe UV light from a young population of stars
are responsible for the extremely red colors. The alternative interpretation
is that these galaxies are star-bursts or active galactic nuclei (AGN),
perhaps triggered by a merging event, In this scenario, the presence
of significant quantities of interstellar dust hides the star-forming
regions or AGN, considerably reddening the observed SEDs. In both
cases, the galaxies are most likely to lie at redshifts $0.8<z<2.0$.
Because EROs appear to be very common objects, both interpretations of the
red colors of EROs have significant implications for our understanding
of galaxy formation and evolution. If the ERO population is dominated
by old ellipticals, then massive galaxy formation was well underway at $z>3$.
Dust-dominated EROs, however, imply that much of the massive galaxy
formation could actually occur at late times, supporting hierarchical galaxy
formation models. To investigate the ERO population further and to decide
which fraction of the EROs are dusty star forming galaxies and which
fraction are maybe old ellipticals, we already begun some additional 
follow-up work. This includes deeper, higher-resolution images and photometry
and additional observations at various wavelengths, e.g. with SCUBA in the
submillimeter.

\section{Final remarks}

In Fig. 13 I present an overview of individual objects and classes
of objects at high redshift we know today (the list of objects
is certainly not complete). Many of the displayed objects
were ``touched'' in this review, others were not, e.g. I did not discuss
high-$z$ radio galaxies and only briefly mentioned the QSOs in connection
with the interesting CADIS result.

Fig. 13 also inlcudes the ``Madau-plot'' (which is cited so often
these days and which I reproduced here from Pettini et al. 1997) 
showing estimates of the star 
formation history in the universe. The filled points are the original points
which Madau (1997) deduced from combining the results of the
complete redshift surveys reaching $z\approx 1$ (e.g. Lilly et al. 1995,
review of Ellis 1997) with the star formation rates and number densities
of the LBGs.
However, the original curve first presented by Madau has been the 
subject of intensive discussion. 
There are several observational
facts which suggest that the points have to be corrected upwards and
that the cosmic star formation rate may not drop as fast at $z>3$.
Note that, as discussed in section 4.2, the LBGs are already dusty. Taking dust
absorption 
into account gives the higher (open) points in Fig. 12
(see Pettini et al. 1997).
Furthermore, the detection of a population of strong Ly-$\alpha $ emitters
out to redshifts of 6 (see chapter 5), which are partly 
missed by the Lyman break technique,
may suggest that the cosmic star formation rates have to be corrected
further upwards at these redshifts (as I indicated by the arrow). 
And last but no least, there might be a 
population of star forming galaxies at high-$z$, which are completely obscured
by dust and therefore not detectable in the optical/near infrared regime.
Such objects could be detected in the far-infrared and sub-millimetre 
due to their dust emission. New instruments like SCUBA on the JCMT
allow for the first time deep submm-surveys of blank sky regions.
First results of such SCUBA surveys (see Hughes et al. 1998, Barger et al.
1998) indeed found significant number densities of submillimetre sources,
which are most likely starbursting galaxies in the redshift range
$2<z<4$. 

The study of high redshift galaxies is a rapidly improving
field of research and the future promises to be even more exiting.
With new instruments like the Next Generation Space Telescope it may 
even be possible to observe supernovae explosions from 
massive star bursts out to $z=10$. 
Theoretical models of galaxy and structure formation will be more and more
constrained by direct observations of the evolution of galaxies starting
from their first emission of star light in the cradle to the grown up systems
they are today.

\acknowledgments
I am grateful to G. Boerner and H. Mo for the invitation to the
workshop, to E. van Kampen and J. Peacock for comments on a draft
of this article and to the Deutsche Forschungsgemeinschaft
for the research fellowship supporting my stay at the Royal Observatory
in Edinburgh.

\newpage
\begin{figure}
\caption{Filter system used by Steidel et al. for
observing the Lyman continuum break at $z\sim 3$ together
with the model spectrum of a young star forming galaxy
taking into account absorption properties of neutral hydrogen
in the galaxy itself and the statistical effects of the
intervening neutral hydrogen (reproduced from Steidel et al. 1998b)} 
\label{fig:UGR_filter}
\end{figure}

\begin{figure}
\caption{Colour evolution with redshift of galaxies of different
spectroscopic type in the $U_n, G, R$ filter systems used by Steidel et al.
to search for Lyman break galaxies (reproduced from Pettini et al. 1997).} 
\label{fig:color_evol}
\end{figure}

\begin{figure}
\caption{These redshift histograms are taken from Steidel et al.(1998b) and
represent the status of Steidel's survey for LBGs in May 1998. In each field
prominent redshift ``spikes'' are clearly visible. The light-grey
histogram indicates the empirical redshift selection function.} 
\label{fig:LBG_hist}
\end{figure}

\begin{figure}
\caption{Spectra of LBG (shifted to the restframe wavelength).
Ly-$\alpha $ appears in emission and absorption. There are
hints of a correlation between the strength of the metallic 
interstellar absorption lines and the strength of Ly-$\alpha $
absorption and of an anti-correlation with Ly-$\alpha $ emission
(from Sprinrad et al. 1998)} 
\label{fig:LBG_spectra_Spin}
\end{figure}

\begin{figure}[h]
\caption{Keck $B$, $R$ and $I$ images of the field around the
radio galaxy 6C0140+326 (z=4.41) together with a detailed 2-d
spectrum showing the Ly-$\alpha $ emission line from the galaxy
at z=5.34 (RD1) (from Dey et al. 1998). BD3 is the $B$-band dropout
Dey et al. originally targeted for.} 
\label{fig:Dey_plate}
\end{figure}

\begin{figure}[h]
\caption{Spectrum of the galaxy RD1 at z=5.34 discovered by Dey et al. (1998).
The right panel shows a magnification of the wavelength region around
the Ly-$\alpha $ line. The solid line corresponds to the z=5.24
galaxy RD1 and the dotted one to the z=4.02 $B$-dropout BD3.
Both lines show the expected asymmetry. (from Dey et al., 1998)} 
\label{fig:Dey}
\end{figure}

\begin{figure}[h]
\caption{Demonstration of the veto-filter strategy. Solid
boxes indicate the band width and 5$\sigma $ limit of the deepest
broad and medium band filters ($B$, $R$, $I_1$, $I_2$ refer to
the scale on the left). The dotted box represents the 5$\sigma $ limit for
line detection in the Fabry-P{\'e}rot scan, and dashed boxes refer to 3$\sigma $
limits in the veto-filters 466/9, 611/18 and 627/15 (scale on the right).
Note that a typical Ly-$\alpha $ galaxy at $z=5.7$ will only be
detected in the Fabry-P{\'e}rot, the $I$-bands and
perhaps the $R$-band, whereas a foreground galaxy
at $z=0.24$ with very strong emission lines should show up in both
466/9, 627/15 and perhaps 611/18.}
\label{fig:veto}
\end{figure}

\begin{figure}[h]
\caption{Photometric spectra from the CADIS data (top panel) together
with Keck spectra (lower panel) of the same object obtained with
the LRIS spectrograph at Keck II and the 1200g/mm grating.
The total exposure time is 8400 s (resolution 4 {\AA} per 
slitwidth). The error bars in wavelength direction of the photometric
CADIS spectra correspond to the transmission wavelengths of the CADIS
filters. According to the signal in the filter 611/16 and lack of signals in 
the filters 465/9 and 628/16, the line in the Fabry-P{\'e}rot was identified 
with the [OIII] emission line of a galaxy at z=0.6277. This classification
is confirmed by the Keck spectrum.} 
\label{fig:nr2787}
\end{figure}

\begin{figure}[h]
\caption{This diagram shows the surface density of the candidates of
Ly-$\alpha $ emitters at z=3.4 reported by Cowie et al. (1998) (points
with error bars) together with
 the limits 
drawn from the CADIS data so far (marked by CADIS in the diagram)
and the limits of the survey for Ly-$\alpha $ emitters at $z_0 = 4.8$ of 
Thompson and Djorgovski
(1995) (marked by TD in the diagram).
We overplot the expectations from a model calculation, which is
consistent with these data points and the limits. 
(Thommes \& Meisenheimer, 1998c).
CADIS searches for Ly-$\alpha $ emitters at redshifts 4.7, 5.7 and 6.55
and should therefore be able to set firm limits on such model calculations.
} 
\label{fig:pgnumber}
\end{figure}

\begin{figure}[h]
\caption{Optical filter set of the CADIS multi-color survey.
They are supplemented by a NIR filter at $\lambda =2100$ nm.
} 
\label{fig:CADIS_filter}
\end{figure}

\begin{figure}[h]
\caption{Comparison between the CADIS number counts and
predictions based on the QSO luminosity functions
(Boyle et al. 1991, WHO=Warren et al. 1994), cumulative
in redshift for $R < 22$ mag. Their is a significant
discrepancy at $z>2.2$, where current literature suggests
counts drop off (for details see Wolf {\it et al.} 1998)
} 
\label{fig:QSO}
\end{figure}

\begin{figure}[h]
\caption{The left panel shows the full colour-magnitude diagram
of the CADIS 16h field. The stepped lines show the mean and $\pm 2\sigma$
colors for the objects. Right panel: Detail of the region which shows the 
extremely red objects (EROs). The filled diamonds are low-mass stars
in our galaxy. The colour of the ultra luminous infrared galaxy
Arp 220 redshifted to z=1.5 is also shown
for comparison (open) circle. The filled circle shows the ERO HR10
(Graham \& Dey, 1996), which is one of the two examples of EROs
studied in detail in the literature. HR10 appears to be a dusty, star-forming
galaxy at z=1.44. 
} 
\label{fig:EROs}
\end{figure}

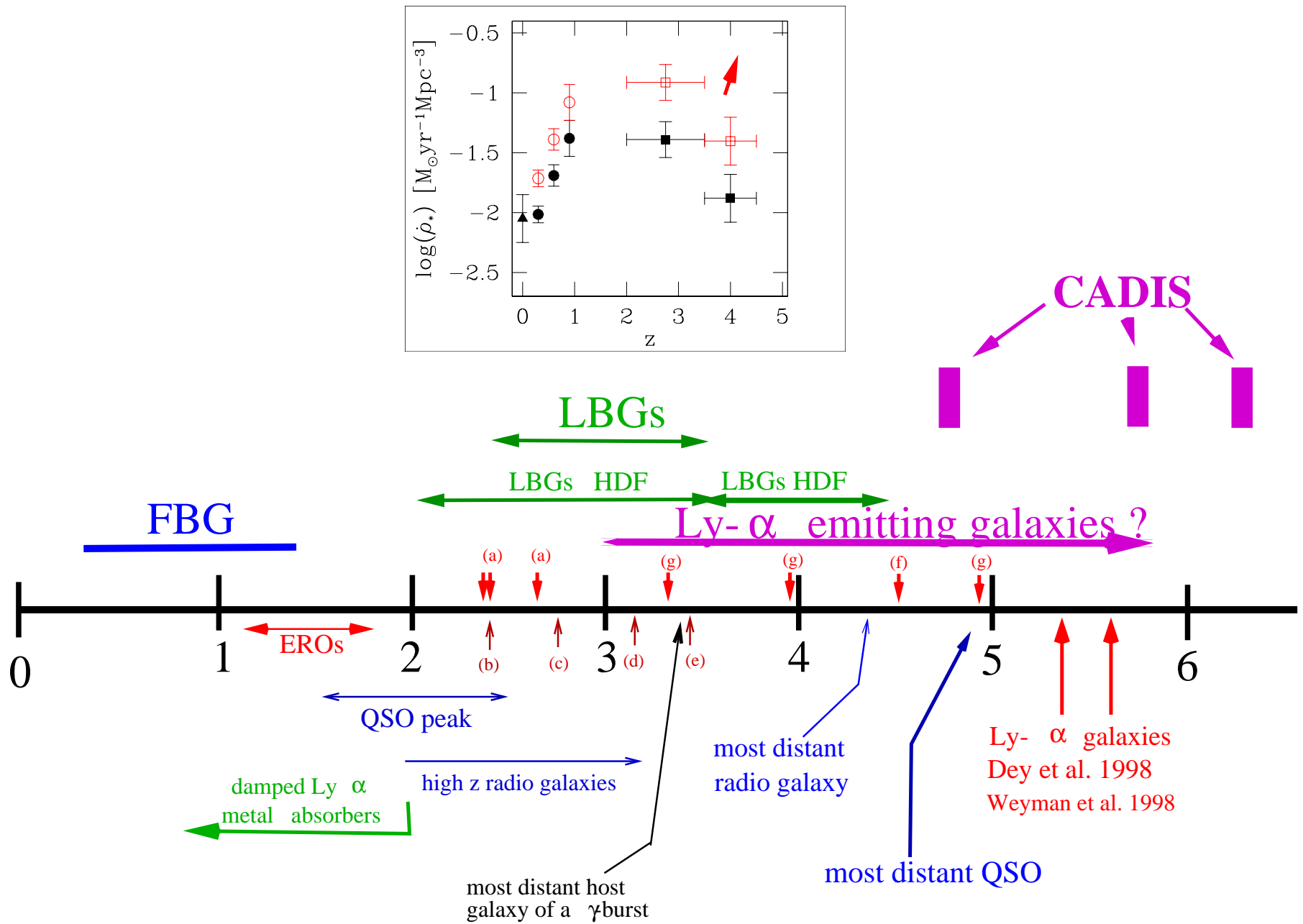
\begin{figure}[h]
\caption{The upper diagram shows the volume-averaged star
formation rate as a function of redshift (Madau 1997, Pettini et al. 1997)
for $H_0=50$kms$^{-1}$. Filled squares: measurements from the HDF;
circles: measurements from the Canada-France Redshift Survey 
(Lilly et al. 1995);
open triangle: from local H$\alpha $ survey; open symbols: same values but
corrected for dust;
{\bf Redshift axis with a selection of high z objects:}
FBG=faint blue galaxies, see review by Ellis (1997);
LBGs = Lyman break galaxies, see section 4;
Ly-$\alpha $ emitting galaxies ? $=>$ see section 5.2;
CADIS $=>$ redshift intervals in which the Calar Alto Deep Imaging
Survey searches for Ly-$\alpha $ emitting galaxies, see section 5.3;
EROs = extremely red objects, see section 5.3.5;
QSO peak $=>$ see section 5.3.5;
most distant radio galaxy, Rawlings et al. 1996;
most distant quasar, Schneider et al. 1991;
(a) emission line objects associated with damped Ly-$\alpha $
and strong metal absorber redshifts, discovered in a narrow-band infrared survey, see Mannucci et al. 1998;
(b), (c), (d), (e) damped Ly-$\alpha $ systems detected in emission, see Macchetto et al. 1993, Djorgovski et al. 1996,
Moller \& Warren 1996, Lowenthal et al. 1991;
(f) Ly-$\alpha $ emitter in the field of a quasar $=>$ see Hu \& McMahon 1996;
(g) Ly-$\alpha$ emitting objects observed behind the rich
cluster CL0939+4713 (Abell 851), see Trager et al. 1997;
(h) sub-galactic, Ly-$\alpha $ emitting clumps with AGN signatures
in the field of a weak radio galaxy, see Pascarelle et al. 1996} 
\label{fig:summary}
\end{figure}

\end{document}